\documentclass[journal]{IEEEtran}

\usepackage{amsmath, amssymb, bm, cite, epsfig, psfrag, mathtools}
\usepackage{epstopdf}
\usepackage{graphicx}
\usepackage{algorithm,algpseudocode}
\usepackage{array}
\usepackage{lipsum}
\usepackage{ragged2e}
\usepackage[font=small]{caption}
\usepackage{bbm}
\usepackage{multirow}
\usepackage[usenames,dvipsnames]{xcolor}
\usepackage{etoolbox}
\usepackage{pbox}
\usepackage{colortbl}
\usepackage{fancyhdr}
\usepackage{tikz}
\usetikzlibrary{calc}

\usepackage{rotating}
\usepackage{dblfloatfix}
\usepackage{supertabular}
\usepackage{longtable}
\usepackage{changepage}
\usepackage{subcaption}
\usepackage{mwe}
\usepackage{color,soul}
\usepackage{booktabs}
\usepackage{capt-of}
\usepackage{footnote}
\makesavenoteenv{tabular}
\makesavenoteenv{table}
\usepackage[hyphens]{url}
\usepackage{cite}
\usepackage{tabu}
\usepackage{fixltx2e}

\usepackage[margin=0.57in]{geometry}
\newcolumntype{P}[1]{>{\centering\hspace{0pt}}p{#1}}
\newcolumntype{M}[1]{>{\centering\hspace{0pt}}m{#1}}
\newcolumntype{L}{>{\centering\arraybackslash}m{3cm}}

\def\dB{\textrm{dB}}

\def\decay{\textrm{decay}}
\def\rise{\textrm{rise}}
\def\mean{\textrm{mean}}

\definecolor{Fgreen}{rgb}{0.13, 0.55, 0.13}

\newtoggle{conference}
\togglefalse{conference} % Use for arxiv version -- also change documentclass
\begin{document}
\bibliographystyle{IEEEtran}

\title{Rapid Fading Due to Human Blockage in Pedestrian Crowds at 5G Millimeter-Wave Frequencies} %\normalsize{\emph{(Invited Paper)}}}
%\vspace{-2cm}
%\author{
%\IEEEauthorblockN{George R. MacCartney, Jr.,~\IEEEmembership{Student Member,~IEEE,} and Theodore S. Rappaport,~\IEEEmembership{Fellow,~IEEE}\\
%\IEEEauthorblockA{NYU WIRELESS\\
%	NYU Tandon School of Engineering, New York University, Brooklyn, NY 11201}\vspace{-0.8cm}
%}
%\thanks{This material is based upon work supported by NOKIA and the NYU WIRELESS Industrial Affiliates Program, three National Science Foundation (NSF) Research Grants: 1320472, 1302336, and 1555332, and the GAANN Fellowship Program. This work is also supported by the National Instruments Lead User Program. The authors thank Y. Xing, J. Koka, R. Wang, and D. Yu for their help in conducting the measurements. G. R. MacCartney, Jr. (email: gmac@nyu.edu) and T. S. Rappaport, are with the NYU WIRELESS Research Center, NYU Tandon School of Engineering, New York University, Brooklyn, NY 11201.}

\author{\IEEEauthorblockN{George R. MacCartney Jr., Theodore S. Rappaport, and Sundeep Rangan}\\
\IEEEauthorblockA{NYU WIRELESS\\
NYU Tandon School of Engineering\\
New York University, Brooklyn, NY 11201\\
\{gmac,tsr,srangan\}@nyu.edu}\vspace{-10mm}

\thanks{This material is based upon work supported by NOKIA and the NYU WIRELESS Industrial Affiliates Program, three National Science Foundation (NSF) Research Grants: 1320472, 1302336, and 1555332, and the GAANN Fellowship Program. The authors thank S. Sun, Y. Xing, H. Yan, J. Koka, R. Wang, and D. Yu, for their help in conducting the measurements.}
}

\maketitle
\begin{tikzpicture}[remember picture, overlay]
\node at ($(current page.north) + (0,-0.25in)$) {G. R. MacCartney, Jr., T. S. Rappaport, and Sundeep Rangan ``Rapid Fading Due to Human Blockage in Pedestrian Crowds };
\node at ($(current page.north) + (0,-0.4in)$) {at 5G Millimeter-Wave Frequencies," \textit{2017 IEEE Global Communications Conference (GLOBECOM)}, Singapore, Dec. 2017.};
\end{tikzpicture}

\begin{abstract}
Rapidly fading channels caused by pedestrians in dense urban environments will have a significant impact on millimeter-wave (mmWave) communications systems that employ electrically-steerable and narrow beamwidth antenna arrays. A peer-to-peer (P2P) measurement campaign was conducted with $7^\circ$, $15^\circ$, and $60^\circ$ half-power beamwidth (HPBW) antenna pairs at 73.5 GHz and with 1 GHz of RF null-to-null bandwidth in a heavily populated open square scenario in Brooklyn, New York, to study blockage events caused by typical pedestrian traffic. \textcolor{black}{Antenna beamwidths that range approximately an order of magnitude were selected to gain knowledge of fading events for antennas with different beamwidths since antenna patterns for mmWave systems will be electronically-adjustable.} Two simple modeling approaches in the literature are introduced to characterize the blockage events by either a two-state Markov model or a four-state piecewise linear modeling approach. Transition probability rates are determined from the measurements and it is shown that average fade durations with a -3 dB threshold are 299.0 ms for $7^\circ$ HPBW antennas and 260.2 ms for $60^\circ$ HPBW antennas. The four-state piecewise linear modeling approach shows that signal strength decay and rise times are asymmetric for blockage events and that mean signal attenuations (average fade depths) are inversely related to antenna HPBW, where $7^\circ$ and $60^\circ$ HPBW antennas resulted in mean signal fades of 15.8 dB and 11.5 dB, respectively. The models presented herein are valuable for extending statistical channel models at mmWave to accurately simulate real-world pedestrian blockage events when designing fifth-generation (5G) wireless systems. 
\end{abstract}

\iftoggle{conference}{}{
}

\section{Introduction}\label{sec:intro}
Millimeter-wave (mmWave) frequencies will play an important role in fifth-generation (5G) communications systems that are predicted to deliver gigabit data throughput speeds to mobile devices~\cite{Rap13a,Boccardi14a}. Large-scale propagation characteristics of mmWave bands have been heavily studied in recent years to generate channel models and simulators for 5G system and network design~\cite{3GPP.38.901,A5GCM15,Sun16b,Sun17b}. However, few outdoor measurements across the mmWave bands have been conducted to understand channel dynamics and fading~\cite{Sun17a,Eliasi15a}. MmWave signals are highly susceptible to blockage which leads to rapid channel variations that have an impact across the protocol stack~\cite{Zhang16b}.

Some indoor studies have been performed for rapid fading caused by human blockage events, particularly at 60 GHz~\cite{Collonge04a,Jacob10b,Jacob09d,Mac16a}. Early human blockage models were developed for the 802.11ad standard which is meant for indoor short-range and high throughput applications~\cite{80211ad10a}. Measurements by Collonge~\textit{et al.} showed that a single person shadowing the direct path between the transmitter (TX) and receiver (RX) would cause more than 20 dB of attenuation for approximately 100 milliseconds (ms), whereas large groups of 11-15 people can cause fade durations of 300 ms or more~\cite{Collonge04a}. Short-range office measurements by Jacob~\textit{et al.} demonstrated that fade durations at 60 GHz in line-of-sight (LOS) environments were on average 550 ms with mean attenuations between 6 dB and 18 dB~\cite{Jacob10b}. 

Outdoor measurements in a street canyon by Weiler~\textit{et al.} using omnidirectional antennas at 60 GHz indicated that attenuation could exceed 40 dB when the LOS path is blocked~\cite{Weiler14a}. In the event of human body shadowing, the LOS multipath component (MPC) experienced extreme fading (more than 20 dB), while secondary reflected MPCs were not affected~\cite{Weiler14a}. An experiment at 28 GHz with omnidirectional antennas in an urban open square studied shadowing caused by pedestrians and vehicles for a small cell, base station-to-mobile scenario~\cite{Weiler16a}. Results indicated 12 dB to 25 dB attenuation of the LOS path for various shadowing events and a double knife-edge diffraction model exhibited a good approximation to human body blockage observations, similar to work in~\cite{Mac16a,Peter12a,Kunisch08a}. 

Work in~\cite{Malik07a} and~\cite{Holtzman94a} showed that fade depths decreased as bandwidth increased, and plateaued at bandwidths of 1 GHz with rapid fading of 4 dB. Similar observations were made in~\cite{Sun17a} at the 73 GHz mmWave band in an urban setting, and spawn the motivation for studying the relationship between antenna beamwidth and signal bandwidth for fading effects caused by human blockage. These studies, however, did not provide a precise statistical characterization of the blockage dynamics that can be used for simulations. This work focuses on rapid fading caused by pedestrians in crowded urban scenarios. We report peer-to-peer (P2P) measurements across a busy walkway in a downtown open square and perform simple statistical techniques to model the results. Section~\ref{sec:measTest} describes the channel sounder and measurement setup and environment. Section~\ref{sec:model} describes two simple models used for analyzing the measurement results. Section~\ref{sec:results} provides the results and analysis and concluding remarks are made in Section~\ref{sec:conc}. 

\begin{figure*}
	\centering
	\includegraphics[trim={0 4.5cm 0 3cm},clip,width=1\textwidth]{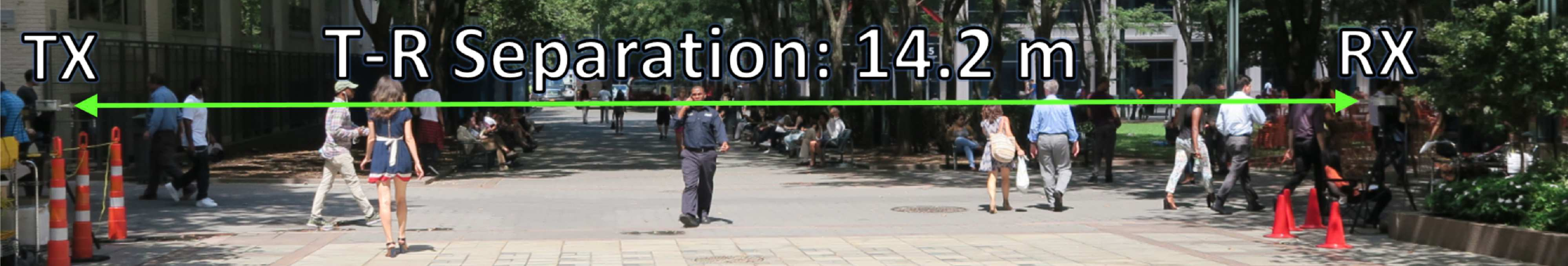}
	\caption{Photo of test environment showing the intersection of Lawrence Street and Myrtle Avenue in downtown Brooklyn, New York.}\label{fig:testPic}
\end{figure*}

\section{Measurement Setup and Hardware}\label{sec:measTest}
\begin{figure}
	\centering
	\includegraphics[trim={0 2cm 0 2.5cm},clip,width=0.48\textwidth]{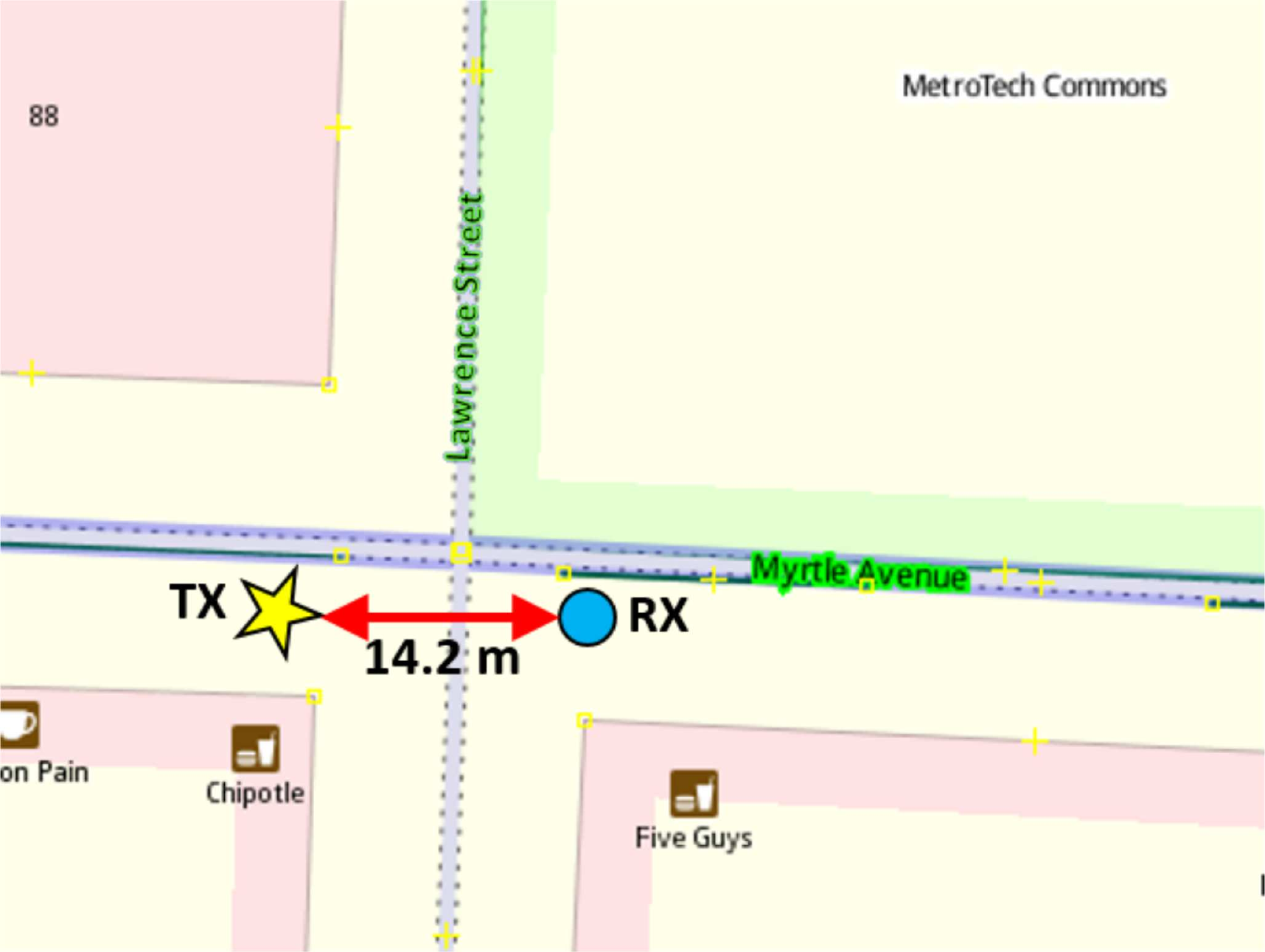}
	\caption{Map of measurement environment in downtown Brooklyn, New York. The yellow star is the TX location and the blue circle is the RX location, with a T-R separation distance of 14.2 m. A majority of traffic moved perpendicular to the line drawn between the TX and RX along Lawrence Street.}\label{fig:map}
	\vspace{0mm}
\end{figure}

The measurements were designed to study the effects of large moving crowds in an urban setting for a peer-to-peer communications links at mmWave. The measurement location is an open square in downtown Brooklyn, New York at the MetroTech Commons courtyard, which is surrounded on all four sides by typical office skyscrapers and the New York University Tandon School of Engineering campus buildings. The measurements were purposely conducted in a high foot-traffic area where pedestrians, child strollers, and bicyclists typically traverse in large quantities. Fig.~\ref{fig:map} shows a map layout of the measurement area where the TX and RX are located on opposite sides of the walkway and were set at heights of 1.5 m relative to the ground, with a transmitter-receiver (T-R) separation distance of 14.2 m. Fig.~\ref{fig:testPic} shows a photo of the test scenario. Table~\ref{tbl:testCase} specifies the number of pedestrians or bicyclists that passed through the link per minute for each test. The pedestrians were never closer than 1.5 m from the TX or RX antennas and on average were equidistant from the TX and RX.

\begin{table}
\centering
\caption{Descriptions of the three measurement tests.}
\label{tbl:testCase}
\begin{center}
	\scalebox{0.95}{
		\begin{tabu}{|c|c|c|c|}
			\hline 
			\textbf{Measurement \#} & \textbf{Test 1}			& \textbf{Test 2}			& \textbf{Test 3}		\\ \specialrule{1.5pt}{0pt}{0pt}
			TX \& RX antenna gain		& 27 dBi				& 20 dBi					& 9.1 dBi				\\ \hline
			TX \& RX Az/El HPBW			& 7$^\circ$/7$^\circ$	& 15$^\circ$/15$^\circ$		& 60$^\circ$/60$^\circ$	\\ \hline
			T-R separation				& \multicolumn{3}{c|}{14.2 m}												\\ \hline
			Observation window			& \multicolumn{3}{c|}{135 s}												\\ \hline
			Total PDPs recorded	& \multicolumn{3}{c|}{40,800}												\\ \hline
			PDP time/frequency interval	& \multicolumn{3}{c|}{$\sim$3.3 ms / 302 Hz}										\\ \hline
			Carrier frequency			& \multicolumn{3}{c|}{73.5 GHz}										\\ \hline
			RF null-to-null bandwidth 	& \multicolumn{3}{c|}{1 GHz}										\\ \hline
			Pedestrian/bicyclist crossings per minute		& 17.8				& 12.9						& 12					\\ \hline
		\end{tabu}}
	\end{center}
	\vspace{-5mm}
\end{table}

Three measurement tests were performed with different pairs of TX and RX horn antennas in order to study the relationship between rapid fading shadowing events and antenna beamwidth. Table~\ref{tbl:testCase} provides the measurement specifications for each test with 7$^\circ$, 15$^\circ$, and 60$^\circ$ azimuth (Az) and elevation (El) HPBW antenna pairs used for test one, test two, and test three, respectively. \textcolor{black}{Three different antenna beamwidth pairs that range approximately an order of magnitude difference were used to gain insights into fading events in pedestrian crossings at mmWave, especially since various phased-array architectures with flexible antenna patterns are expected for base stations and mobile handsets. It is also expected that the largest antenna beamwidth for mmWave systems will be no more than 60$^\circ$ which is why it was the maximum beamwidth used for this experiment~\cite{Sun17a}.} For each test, the antennas were boresight-aligned and remained fixed. Each test consisted of a 135-second free-running observation window in which power delay profiles (PDPs) were captured with a $\sim$302 Hz snapshot interval. Each observation window resulted in 40,800 PDP snapshots with a $\sim$3.3 ms time separation between back-to-back PDPs. A -20 dB max peak threshold was applied to each PDP in post-processing and the area under each of the PDPs was integrated to calculate the total received power for every snapshot. We note that due to spatial filtering by the directional antennas, only one multipath cluster (LOS path) was detectable over all measurements. 

The real-time spread spectrum correlator (direct-correlation) channel sounder described in~\cite{Mac17a} was used for the measurements at a 73.5 GHz carrier frequency with 1 GHz of RF null-to-null bandwidth and a 2 ns MPC time resolution. True propagation delay timing was calibrated as described in~\cite{Mac17a}. The maximum effective isotropic radiated power (EIRP) was 21.2 dBm when the channel sounder was configured with a TX output power of -5.8 dBm and with the 27 dBi gain and 7$^\circ$ Az/El HPBW TX antenna. Additional details for the channel sounder are provided in~\cite{Mac17a}.

\section{Modeling Methodology}\label{sec:model}
Of many potential modeling strategies, two simple methodologies are used here to model the blockage events: 1) a two-state approach with unshadowed and shadowed states; 2) a four-state approach with unshadowed, decaying signal level, shadowed, and rising signal level states. 

\subsection{Two-State Blockage Modeling}\label{sec:modelTwo}
\begin{figure}
	\centering
	\includegraphics[width=0.41\textwidth]{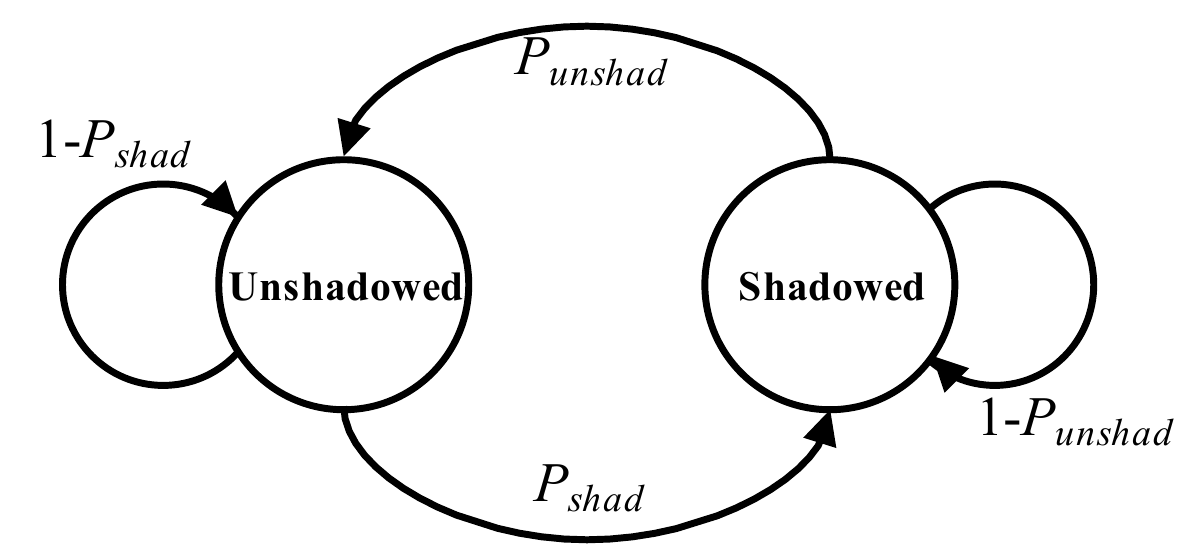}
	\caption{Two-state Markov model for blockage events.}\label{fig:twoStateMarkov}
\end{figure}

\begin{figure}
	\centering
	\includegraphics[width=0.48\textwidth]{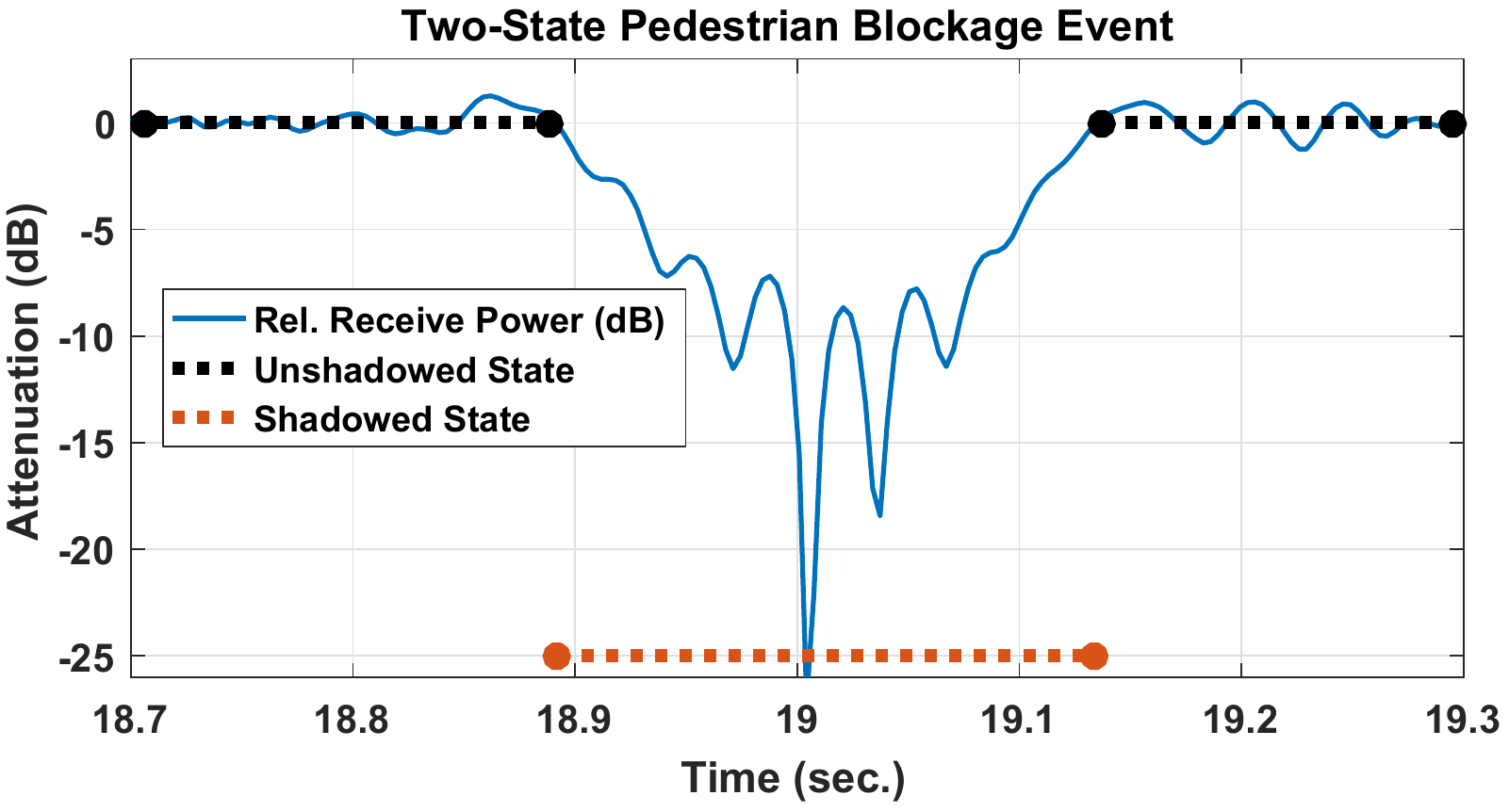}
	\caption{Two-state shadowing event with 0 dB threshold showing unshadowed (black line) and shadowed (red line) regions}\label{fig:twoStateData}
	\vspace{-5mm}
\end{figure}

A simple two-state Markov model can be used to characterize unshadowed and shadowed states for a wireless link in the presence of pedestrian induced variations in received signal strength~\cite{Kashiwagi10a,Dehnie07a}. Fig.~\ref{fig:twoStateMarkov} shows a diagram of a two-state Markov model where $P_{unshad}$ and $P_{shad}$ indicate the transition probabilities of going from a shadowed to unshadowed state and an unshadowed to shadowed state, respectively. A predefined threshold is selected in order to characterize a shadowed or unshadowed state and is typically 0 dB to -10 dB from the zero-crossings just before and after the fading event. Fig.~\ref{fig:twoStateData} depicts the characterization of a typical blockage event with two-states when applying a 0 dB threshold relative to the zero-crossings for the beginning and end of a shadowing event.

\subsection{Four-State Blockage Modeling}\label{sec:modelFour}
\begin{figure}[t!]
	\centering
	\includegraphics[width=0.46\textwidth]{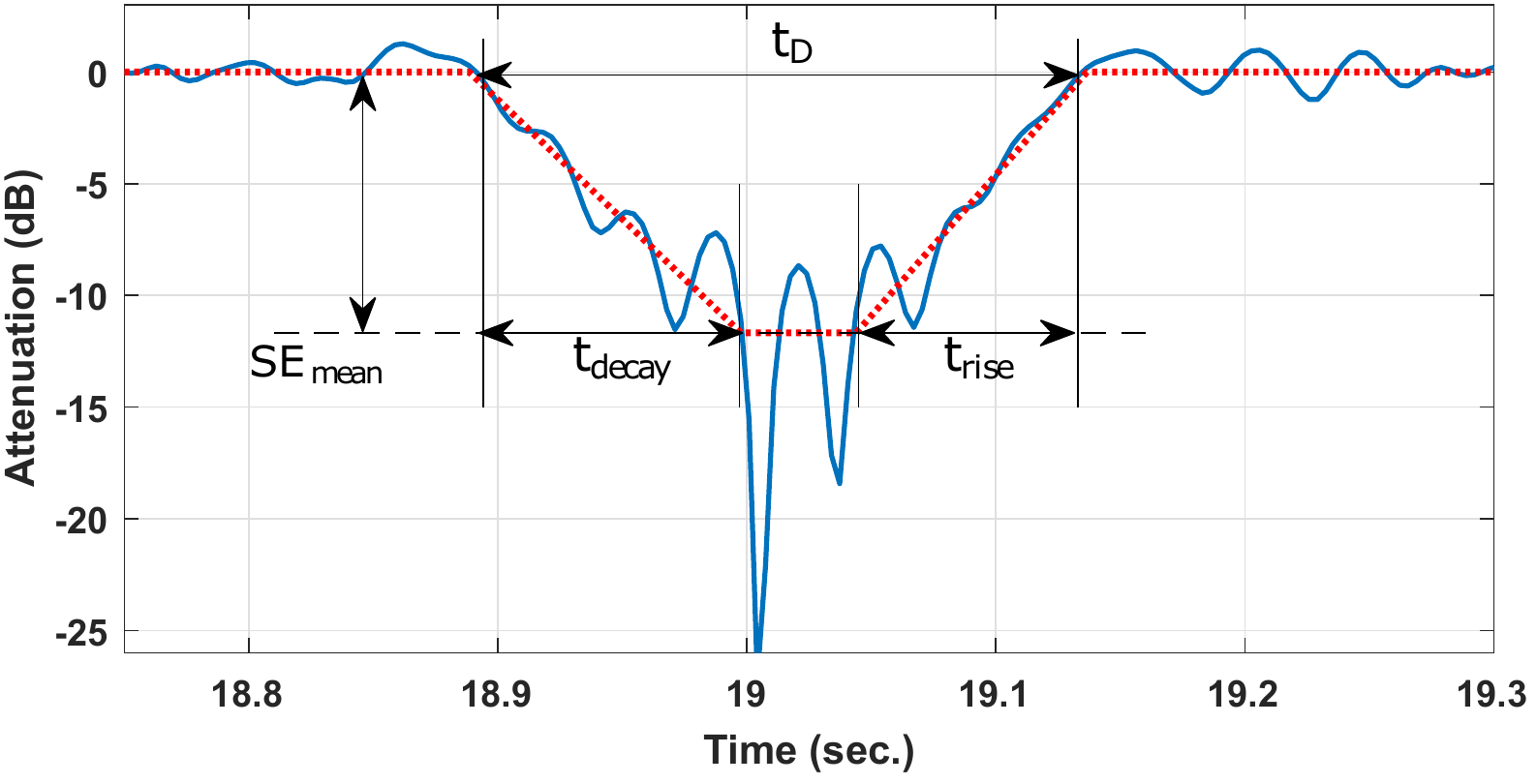}
	\caption{Four-state shadowing event with an $SE_{\mean}$ calculated threshold and labels for decay time $t_{\decay}$, rise time $t_{\rise}$, shadowing event mean attenuation $SE_{\mean}$, and fade duration $t_D$.}\label{fig:fourStateData}
\end{figure}

A four-state piecewise linear modeling approach is also useful for characterizing blockage with the following regions: unshadowed, shadowed, a decaying signal level region from unshadowed to shadowed, and a rising signal level region from shadowed to unshadowed~\cite{Jacob10b}. The blockage event signal level (in dB) as a function of time, $SE(t)$ is represented as~\cite{Jacob10b,Peter12a}:
\begin{equation}
\footnotesize
SE(t)[\dB] = \begin{cases}
r_{\decay}\cdot t,	&\text{for } 0\leq t \leq\frac{SE_{\mean}}{r_{\decay}}\\
SE_{\mean}, 			&\text{for }\frac{SE_{\mean}}{r_{\decay}}\leq t \leq t_D-\frac{SE_{\mean}}{r_{\rise}}\\
SE_{\mean}-r_{\rise}\cdot t,	&\text{for } t_D-\frac{SE_{\mean}}{r_{\rise}} \leq t \leq t_D\\
0,	& \text{otherwise}
\end{cases}
\end{equation}
where $r_{\decay}$ is the signal strength decay rate in dB/ms, $t$ is the time in ms from the onset ($t=0$ ms) of the blockage event, $SE_{\mean}$ is the blockage event mean signal attenuation in dB, $r_{\rise}$ is the blockage event signal strength rise rate in dB/ms, and $t_D$ is the shadowing event fade time duration in ms~\cite{Jacob10a}. The $r_{\decay}$ and $r_{\rise}$ rates are determined by the decay time $t_{\decay}$, rise time $t_{\rise}$, and either a predetermined threshold between the last zero-crossing before and the first zero-crossing after the rapid signal fade, or is strictly based on the mean signal attenuation of the shadowing event $SE_{\mean}$, calculated over the interval [$\frac{1}{3}t_D < t < \frac{2}{3}t_D$]. Regardless of the threshold method used, the blockage event mean signal attenuation $SE_{\mean}$ is determined over the same interval. Fig.~\ref{fig:fourStateData} shows an example of a blockage event with a threshold of $SE_{\mean}$. 

\begin{figure}[t!]
	\centering
	\includegraphics[width=0.46\textwidth]{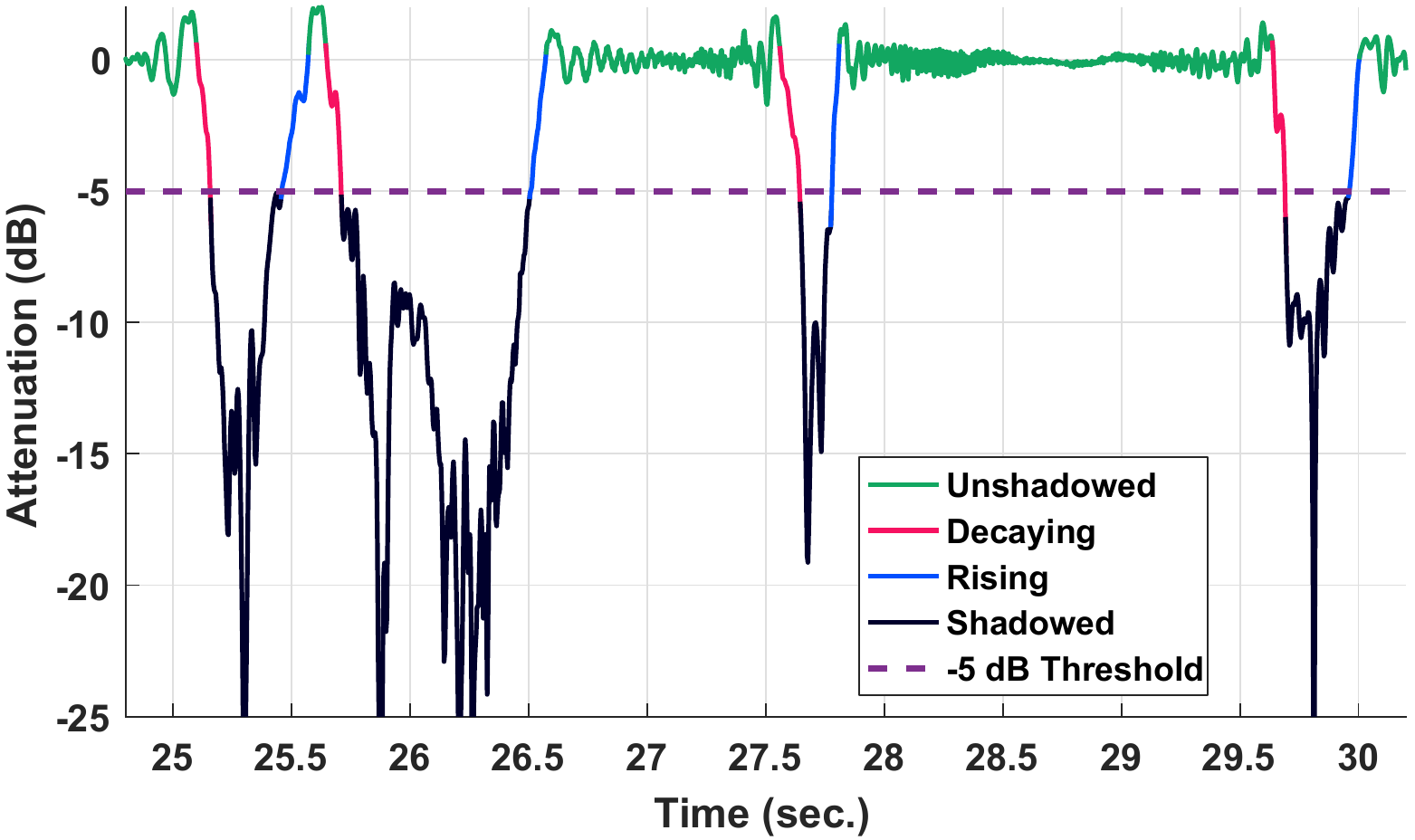}
	\caption{Four-state piecewise linear model with unshadowed, decaying, shadowed, and rising regions during a 73.5 GHz blockage event with a -5 dB pre-defined threshold, as an example.}\label{fig:fourStateEx}
\end{figure}
\begin{figure}[t!]
	\centering
	\includegraphics[width=0.42\textwidth]{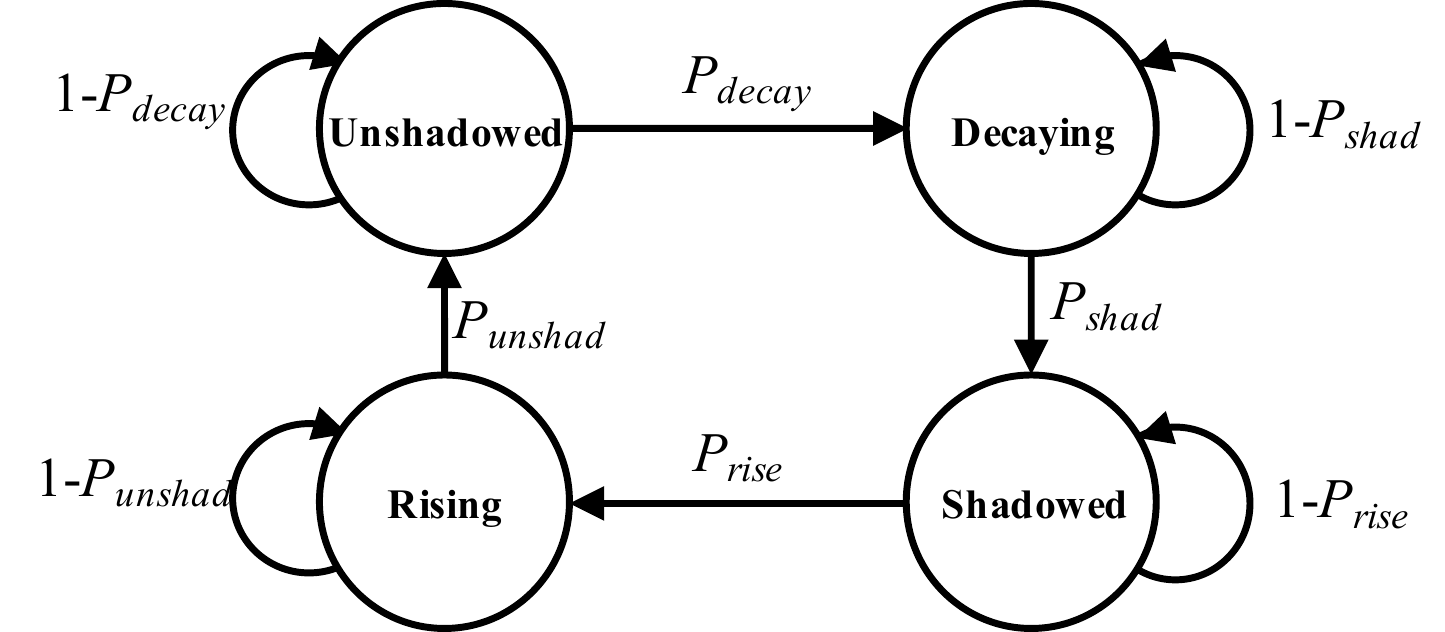}
	\caption{Four-state Markov model for unshadowed, decaying, shadowed, and rising regions for modeling blockage events.}\label{fig:fourStateMarkov}
\end{figure}

The selected threshold level will affect the decay and rise times of the blockage events, and by consequence, will influence the decay and rise rates of signal strength. Therefore, blockage modeling for different applications may require using various thresholds depending upon the system requirements and fade margin limitations. For example, the blockage events in Fig.~\ref{fig:fourStateEx} were determined with a -5 dB threshold and the unshadowed, decaying, shadowed, and rising regions are indicated by different colors. 

The example in Fig.~\ref{fig:fourStateEx} brings about a four-state Markov model shown in the diagram in Fig.~\ref{fig:fourStateMarkov}. The transition probabilities in Fig.~\ref{fig:fourStateMarkov} are defined as: $P_{decay}$ is the probability of transitioning from an unshadowed state to a state of decaying signal levels, $P_{shad}$ is the probability of transitioning from a state of decaying signal levels to a shadowed state,  $P_{rise}$ is the probability of transitioning from a shadowed state to a rising signal level state, and $P_{unshad}$ is the probability of transitioning from a rising signal state to an unshadowed/unblocked state. 

%More complicated modeling techniques such as the forward-backward algorithm to determine hidden Markov models or the alternating direction method of multipliers (ADMM) algorithm may be used to characterize the blockage event states. Although, the simple models described herein are more than sufficient for modeling human blockage events.  

\section{Measurement Results and Analysis}\label{sec:results}
\subsection{Two-State Blockage Model}\label{sec:resultsTwo}
\textcolor{black}{While various thresholds could be selected to define the start of a shadowing event, a -3 dB threshold from the unshadowed 0 dB attenuation level was used to compare the blockage observations for different antenna pairs. Note that blockage events were initially labeled using the Lloyd-Max algorithm (2 states) in MATLAB, but were analyzed and modeled using the criteria and threshold defined here. Table~\ref{tbl:twoStateProbs} provides the transition probability rate $\lambda$ values of conditional state transitions for the observed blockage events. For a transition probability $p$ and a sampling interval $T$, $\lambda = p/T$. For example, the probability of transitioning from a shadowed state to an unshadowed state with 7$^\circ$ HPBW antennas is $p=0.0007$ for the sampling interval $T=3.3$ ms, such that $\lambda=0.21\text{ transitions/second}$ ($\text{trans/sec}$). In Table~\ref{tbl:twoStateProbs}, $\lambda$ represents the conditional probability rate for transitioning from a ``current state" to the ``next state."}

\textcolor{black}{The transition rates for a shadowed state conditioned on an unshadowed state are consistent and around 0.20 trans/sec for the three antenna HPBW pairs and for all antenna tests combined. These transition rates imply that there are approximately 5 seconds between shadowed states. The transition rates are slightly different for an unshadowed state conditioned on a shadowed state, where the rate is correlated with the antenna HPBW. For instance, $\lambda$ is 3.36 and 3.85 for $7^\circ$ and $60^\circ$ HPBW TX/RX antennas, respectively, indicating that blockage events are longer for narrower beamwidth antennas ($7^\circ$: 298 ms) compared with wider beamwidth antennas ($60^\circ$: 260 ms). This intuitively makes sense since wider beamwidth antennas would result in more energy spread around the obstructions from the transmitter, and a wider viewing angle at the receiver to capture diffracted energy around the pedestrian blockers.}

\begin{figure}[t!]
	\centering
	\includegraphics[width=0.45\textwidth]{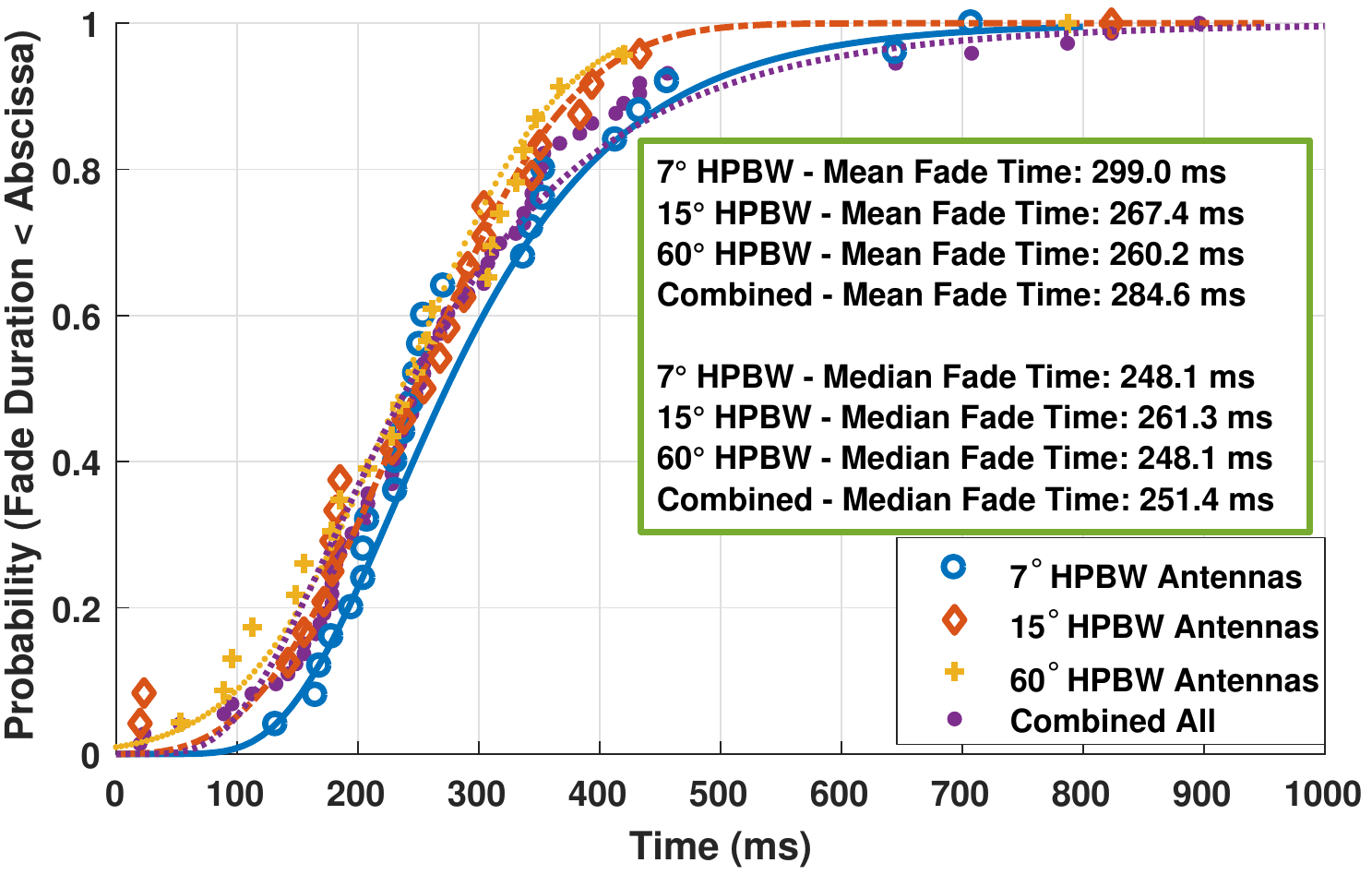}
	\caption{Two-state Markov model mean fade time duration (ms) CDFs for the 73.5 GHz blockage measurements. The lines in the plot represent the CDF curves best-fit to the data with parameters provided in Table~\ref{tbl:twoStateCDFs}.}\label{fig:twoStateMarkovFadeDurationCDF}
\end{figure}

Fig.~\ref{fig:twoStateMarkovFadeDurationCDF} shows the cumulative distribution function (CDF) curves of the pedestrian blockage event fade time durations for each set of antennas and \textcolor{black}{ when blockage event observations from all antenna sets are considered together for analysis}. Table~\ref{tbl:twoStateCDFs} provides the best-fit CDF parameters\textsuperscript{*} to the empirical CDF data in Fig.~\ref{fig:twoStateMarkovFadeDurationCDF} (using the \emph{fitdist} function in MATLAB\textsuperscript{\textregistered}) and their goodness of fit (GOF) to the data (using the \emph{goodnessOfFit} function in MATLAB\textsuperscript{\textregistered}) where the normalized mean square error (NMSE) GOF measure ranges from $-\infty$ to 1, where $-\infty$ indicates a poor fit, and 1 indicates a perfect fit~\cite{Rap15b}. Similar to the transition probability rates, the mean fade time duration reduces as antenna beamwidth increases. For example, the mean fade time durations are 299.0 ms, 267.4 ms, and 260.2 ms for $7^\circ$, $15^\circ$, and $60^\circ$ TX/RX HPBW antenna pairs, respectively. The GOF values for the CDFs in Table~\ref{tbl:twoStateCDFs} are all 0.95 or above and indicate a good match to the empirical data. 

\begin{table}[b!]
	\centering
	\caption{Two-state Markov model transition probability rates for human blockage events at 73.5 GHz.}
	\label{tbl:twoStateProbs}
	\begin{center}
		\scalebox{0.90}{
			\begin{tabu}{|c|c|c|c|} \hline
				TX/RX HPBW 						& Current State		& Next State	&	$\lambda$ - rate (trans/sec) \\ \specialrule{1.5pt}{0pt}{0pt}
				\multirow{2}{*}{$7^\circ$} 		& unshadowed 		& shadowed		&	0.21 \\ \cline{2-4}
				& shadowed 			& unshadowed	&	3.36 \\ \hline
				\multirow{2}{*}{$15^\circ$} 	& unshadowed 		& shadowed		&	0.21 \\ \cline{2-4}
				& shadowed 			& unshadowed	&	3.42 \\ \hline
				\multirow{2}{*}{$60^\circ$} 	& unshadowed 		& shadowed		&	0.18 \\ \cline{2-4}
				& shadowed 			& unshadowed	&	3.85 \\ \hline
				\multirow{2}{*}{All Combined} 	& unshadowed 		& shadowed		&	0.18 \\ \cline{2-4}
				& shadowed 			& unshadowed	&	3.52 \\ \hline
				%				\midrule
				%				\textbf{7$^\circ$ Az/El HPBW Antennas} & unshadowed & shadowed \\ \midrule
				%				unshadowed & 0.9993 & 0.0007 \\ 
				%				shadowed & 0.0111 & 0.9889 \\ \midrule
				%				\textbf{15$^\circ$ Az/El HPBW Antennas} & unshadowed & shadowed \\ \midrule
				%				unshadowed & 0.9993 & 0.0007 \\ 
				%				shadowed & 0.0113 & 0.9887 \\ \midrule 
				%				\textbf{60$^\circ$ Az/El HPBW Antennas} & unshadowed & shadowed \\ \midrule
				%				unshadowed & 0.9994 & 0.0006 \\ 
				%				shadowed & 0.0127 & 0.9873 \\ \midrule
				%				\textbf{All Antenna Combined}& unshadowed & shadowed \\ \midrule
				%				unshadowed & 0.9994 & 0.0006 \\ 
				%				shadowed & 0.0116 & 0.9884 \\ \midrule
			\end{tabu}
		}
	\end{center}
\end{table}
\begin{table}[t!]
	\centering
	\caption{CDF parameters for the two-state Markov model.}
	\label{tbl:twoStateCDFs}
	\begin{center}
		\scalebox{0.92}{
			\begin{tabu}{|c|c|c|c|}
				\hline 
				\multicolumn{4}{|c|}{\textbf{CDF Parameters for Fade Duration [ms] of Two-State Markov Model}}  \\ \hline
				\textbf{Ant. HPBW} & \textbf{Dist.} & \textbf{Parameters} & \textbf{GOF}  \\ \specialrule{1.5pt}{0pt}{0pt}
				7$^\circ$ & Log-normal & $\mu = 5.61$; $\sigma = 0.42$ & 0.95	\\ \hline
				15$^\circ$ & Weibull & $\alpha = 282.12$; $\beta = 2.84$ & 0.98	\\ \hline
				60$^\circ$ & Normal & $\mu = 236.22$; $\sigma = 100.50$ & 0.98	\\ \hline
				7$^\circ$; 15$^\circ$; 60$^\circ$ & Log-normal & $\mu = 5.48$; $\sigma = 0.54$ & 0.98	\\ \hline
			\end{tabu}
		}
	\end{center}\vspace{-5mm}
\end{table}

While 100's of milliseconds for a fade duration may seem short, it is significant for wireless communications systems. For instance, the 802.11ad standard for 60 GHz has packet transmission times on the microsecond level and thus a blockage event of 200 ms to 300 ms would result in severe throughput degradation or an outage over that time span. Therefore, access point diversity with multiple nodes can be coordinated in MAC protocols to maintain a reliable connection to a mobile or user equipment (UE)~\cite{Singh09a}. Additionally, electrically-steerable antennas envisioned for mmWave communications can be used to beam switch around obstacles in order to find reflections and scatterers that avoid pedestrian blockages. 

\subsection{Four-State Blockage Model}\label{sec:resultsFour}
The piecewise linear model described in Section~\ref{sec:modelFour} was used to model four states of blockage events: unshadowed, decaying signal strength, shadowed, and rising signal strength. Similar to the two-state model in Section~\ref{sec:resultsTwo}, blockage events were initially labeled using the Lloyd-Max algorithm in MATLAB, but were analyzed and modeled using the criteria and threshold defined here. For the results shown here, a 0 dB threshold for the first and last zero-crossings of the blockage event was used to determine the beginning and end times of the decaying and rising states, respectively~\cite{Jacob10b,Peter12a}. The mean attenuation level calculated over the interval [$\frac{1}{3}t_D < t_D < \frac{2}{3}t_D$] was used as the threshold for determining the shadowed state. Following this procedure, the shadowing event fade time durations $t_D$, decaying signal strength rates $r_{\decay}$, the mean attenuation of shadowing events $SE_{\mean}$, and rising signal strength rates $r_{\rise}$ were determined for each individual blockage event of the measurement tests.

\begin{figure}
	\centering
	\begin{subfigure}[b]{0.46\textwidth}
		\centering
		\includegraphics[width=\textwidth]{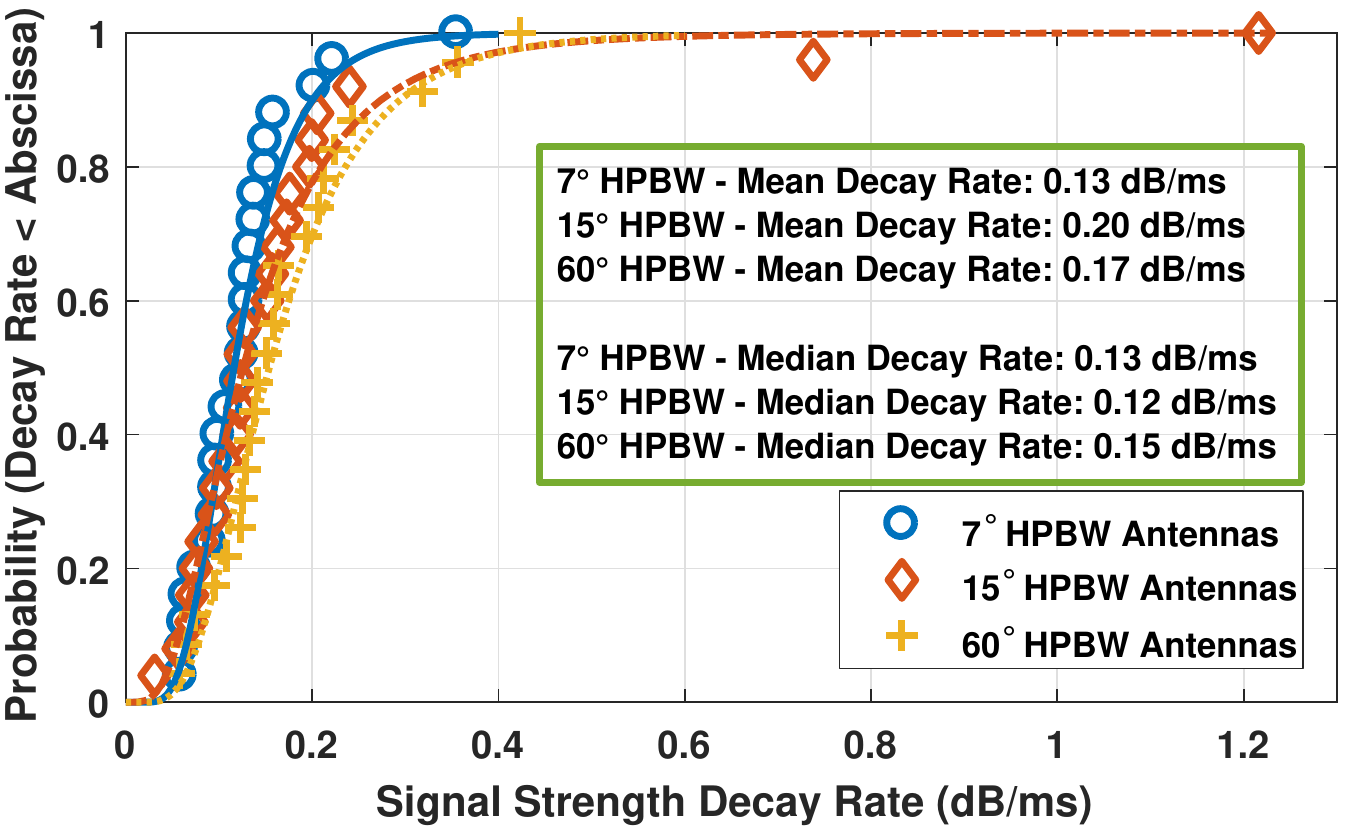}
		\caption{Signal strength decay rate $r_{\decay}$ (dB/ms) CDFs}   
		\label{fig:fourStateBlockDecayRateCDF}
	\end{subfigure}
	\hfill
	\begin{subfigure}[b]{0.46\textwidth}  
		\centering 
		\includegraphics[width=\textwidth]{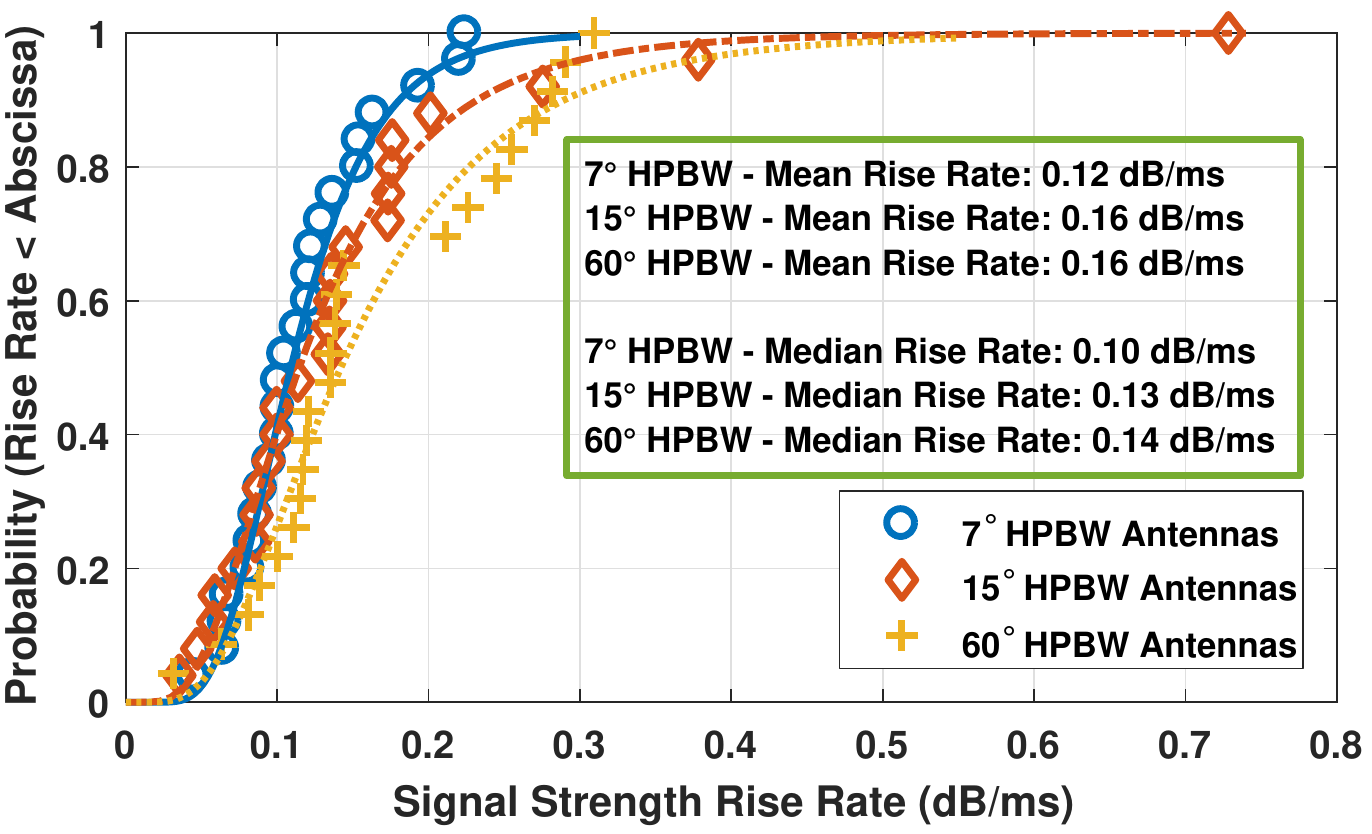}
		\caption{Signal strength rise rate $r_{\rise}$ (dB/ms) CDFs}      
		\label{fig:fourStateBlockRiseRateCDF}
	\end{subfigure}
	\hfill
	\begin{subfigure}[b]{0.46\textwidth}   
		\centering 
		\includegraphics[width=\textwidth]{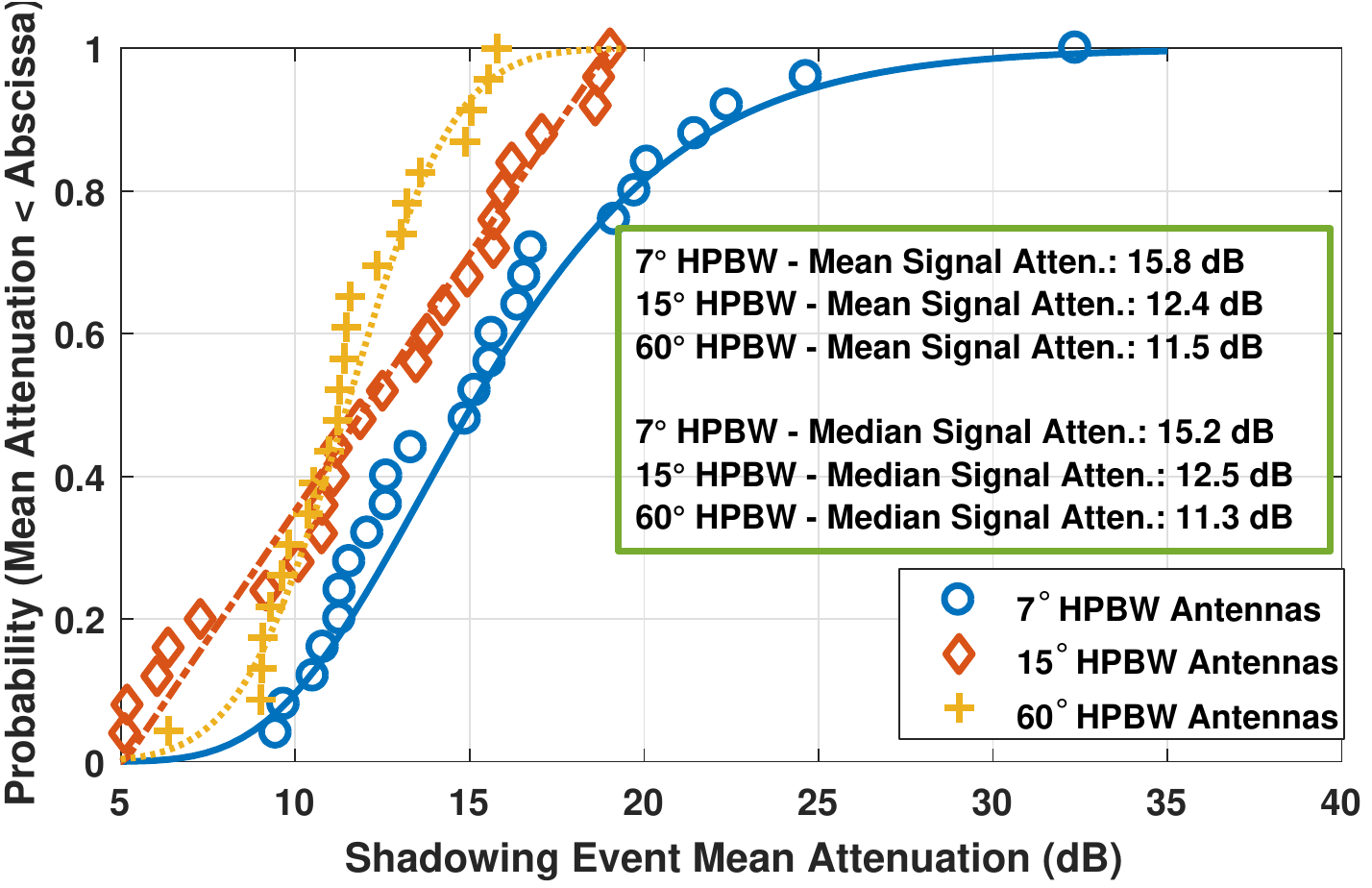}
		\caption{Shadowing event mean attenuation $SE_{\mean}$ (dB) CDFs}     
		\label{fig:fourStateBlockMeanAttenCDF}
	\end{subfigure}
	\hfill
	\begin{subfigure}[b]{0.46\textwidth}   
		\centering 
		\includegraphics[width=\textwidth]{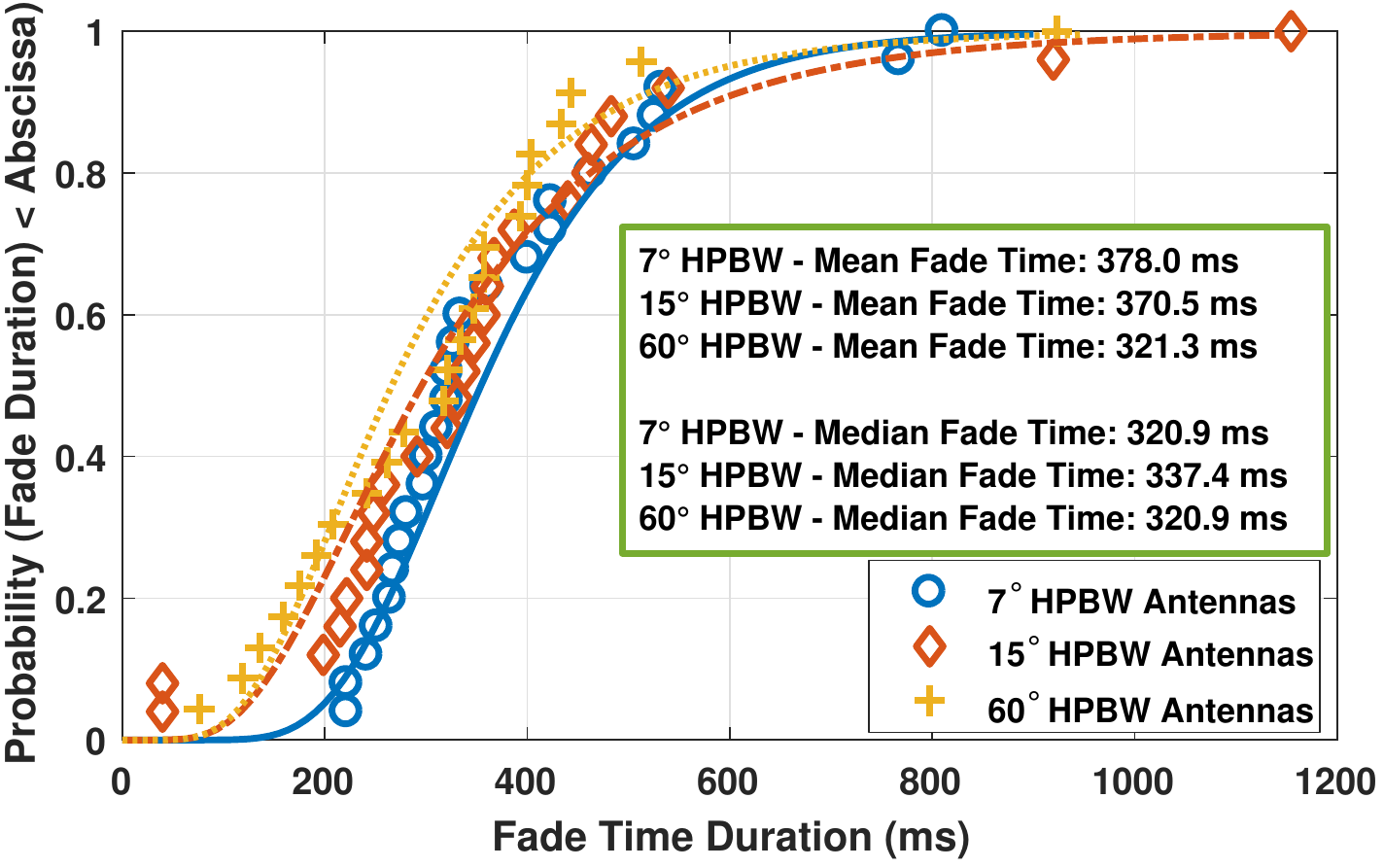}
		\caption{Fade time duration $t_D$ (ms) CDFs}    
		\label{fig:fourStateBlockTimeCDF}
	\end{subfigure}
	\caption{Four-state piecewise linear blockage model CDFs for decay rate, rise rate, mean signal attenuation, and fade time duration for each pair of HPBW antennas at 73.5 GHz. \textcolor{black}{The lines in (a) to (d) represent the CDF curves best-fit to the data with parameters provided in Table~\ref{tbl:fourStateCDFs}}.} 
	\label{fig:fourStateCDFs}
\end{figure}

Fig.~\ref{fig:fourStateBlockDecayRateCDF} shows the CDFs of the signal strength decay rates in dB/ms for the three TX/RX antenna HPBW pairs along with the best-fit CDF curves\footnote{Note that all best-fit CDFs were derived using the \emph{fitdist} function and all GOF values were calculated using the \emph{goodnessOfFit} function, both in MATLAB\textsuperscript{\textregistered}. We also note that due to the limited measured data, the last few percent of data of each of the tails was ignored when solving for the best-fit distributions.} to the empirical data and with best-fit model parameters provided in Table~\ref{tbl:fourStateCDFs}. The GOF values are all 0.93 or above, indicating a good distribution fit to the data. The signal strength decay rates are determined as the rate of change over the time-frame ($t_{\decay}$) from the last zero-crossing before the fading event and the first instance of the signal level at or below the mean attenuation value. The median values for the signal strength decay rates give a better indication of the trend than the mean values since the 15$^\circ$ HPBW antenna tests had two extremely rapid decay rates of 0.74 dB/ms and 1.22 dB/ms when a hand waved two times between the LOS link, which skewed the mean decay rate. The median decay rates of 0.13 dB/ms and 0.12 dB/ms for the 7$^\circ$ and 15$^\circ$ HPBW antennas, respectively, show that relatively narrowbeam TX and RX antennas result in similar blockage decay rates. However, the median decay rate of 0.15 dB/ms for the 60$^\circ$ HPBW antennas indicates a much sharper drop for wider beamwidth antennas. Initially, this may not seem intuitive, but since the 60$^\circ$ HPBW sectored antennas have a wider viewing angle, their mean attenuations may not be as significant as for narrower beamwidth antennas such that a drop to the mean signal attenuation occurs at a faster rate. Additional studies are needed to support this conjecture. 

\begin{table}[b!]
	\centering
	\caption{Best-fit CDF parameters of the Four-State Markov models for the 73.5 GHz rapid fading events caused by pedestrian traffic.}
	\label{tbl:fourStateCDFs}
	\begin{center}
		\scalebox{0.95}{
			\begin{tabu}{|c|c|c|c|}
				\specialrule{1.5pt}{0pt}{0pt}
				\multicolumn{4}{|c|}{\textbf{CDF Parameters for Signal Strength Decay Rate: $\bm{r_{\decay}}$ [dB/ms]}}  \\ \hline
				\textbf{Ant. HPBW} & \textbf{Dist.} & \textbf{Parameters} & \textbf{GOF}  \\ \hline
				7$^\circ$ & Log-normal & $\mu = -2.15$; $\sigma = 0.43$ & 0.96	\\ \hline
				15$^\circ$ & Log-normal & $\mu = -2.07$; $\sigma = 0.61$ & 0.98	\\ \hline
				60$^\circ$ & Log-normal & $\mu = -1.87$; $\sigma = 0.51$ & 0.98	\\ \specialrule{1.5pt}{0pt}{0pt}
				\multicolumn{4}{|c|}{\textbf{CDF Parameters for Signal Strength Rise Rate: $\bm{r_{\rise}}$ [dB/ms]}}  \\ \hline
				\textbf{Ant. HPBW} & \textbf{Distribution} & \textbf{Parameters} & \textbf{GOF}  \\ \hline
				7$^\circ$ & Log-normal & $\mu = -2.22$; $\sigma = 0.39$ & 0.99	\\ \hline
				15$^\circ$ & Log-normal & $\mu = -2.17$; $\sigma = 0.55$ & 0.99	\\ \hline
				60$^\circ$ & Log-normal & $\mu = -1.95$; $\sigma = 0.56$ & 0.95	\\ \specialrule{1.5pt}{0pt}{0pt}
				\multicolumn{4}{|c|}{\textbf{CDF Parameters for Shadowing Event Mean Attenuation: $\bm{SE_{\mean}}$ [dB]}}  \\ \hline
				\textbf{Ant. HPBW} & \textbf{Distribution} & \textbf{Parameters} & \textbf{GOF}  \\ \hline
				7$^\circ$ & Log-normal & $\mu = 2.71$; $\sigma = 0.31$ & 0.97	\\ \hline
				15$^\circ$ & Uniform & $A = 5.11$; $B = 19.02$ & 0.98	\\ \hline
				60$^\circ$ & Normal & $\mu = 11.50$; $\sigma = 2.41$ & 0.96	\\ \specialrule{1.5pt}{0pt}{0pt}
				\multicolumn{4}{|c|}{\textbf{CDF Parameters for Fade Time Duration: $\bm{t_{D}}$ [ms]}}  \\ \hline
				\textbf{Ant. HPBW} & \textbf{Distribution} & \textbf{Parameters} & \textbf{GOF}  \\ \hline
				7$^\circ$ & Log-normal & $\mu = 5.87$; $\sigma = 0.35$ & 0.94	\\ \hline
				15$^\circ$ & Log-normal & $\mu = 5.69$; $\sigma = 0.53$ & 0.93	\\ \hline
				60$^\circ$ & Log-normal & $\mu = 5.58$; $\sigma = 0.49$ & 0.95	\\ \specialrule{1.5pt}{0pt}{0pt}
		\end{tabu}}
	\end{center}
\end{table}

The signal strength rise time ($r_{\rise}$) CDFs in Fig.~\ref{fig:fourStateBlockRiseRateCDF} show that wider beamwidth antennas result in faster rise time rates than narrowbeam antennas. For instance, the 15$^\circ$ and 60$^\circ$ HPBW TX/RX antenna pairs resulted in 0.13 dB/ms and 0.14 dB/ms median rising signal strength rates compared to a 0.10 dB/ms increasing signal strength rate for the 7$^\circ$ HPBW TX/RX antennas. A noteworthy observation is that the decay ($r_{\decay}$) and rise ($r_{\rise}$) time rates are asymmetric, which was also noticed in~\cite{Peter12a}. The exact reason for this is unknown, but perhaps is an effect of the trajectories with which blockers cross the LOS path between the TX and RX antennas. Table~\ref{tbl:fourStateCDFs} gives the CDF parameters for the best-fit curves to empirical data in Fig.~\ref{fig:fourStateBlockRiseRateCDF} along with high GOF values.

The mean signal attenuation for the three blockage tests resulted in an expected trend where the narrower beamwidth antennas demonstrated greater mean signal attenuations for individual blockage events, as shown by the CDFs in Fig.~\ref{fig:fourStateBlockMeanAttenCDF} (See Table~\ref{tbl:fourStateCDFs} for parameters). Specifically, the 7$^\circ$ HPBW antennas produced an average signal attenuation of 15.8 dB whereas the 15$^\circ$ and 60$^\circ$ HPBW antennas demonstrated an average signal fade of 12.4 dB and 11.5 dB, respectively. The inverse relationship between antenna HPBW and mean signal attenuation can be attributed to the larger beamwidth antennas capturing more diffracted energy around the human blockers during the blockage event, compared to narrower beam antennas that act as spatial filters.

\textcolor{black}{A simple analytical expression may be used to determine the mean signal attenuation as a function of the TX/RX antenna HPBW:
\begin{equation}\label{eq:modelFit}
\footnotesize
\text{Mean Blockage Attenuation [dB]}=10\log_{10}\left(b+\frac{180}{A_{BW}}\right)
\end{equation}
where $b$ is a constant and $A_{BW}$ is the TX/RX antenna HPBW in degrees. Fig.~\ref{fig:BlockModelFit} shows all mean attenuations for each TX/RX HPBW antenna pair and the average mean attenuation for each pair. A rough estimate for the model in~\eqref{eq:modelFit} is fit to the average mean signal attenuations in Fig.~\ref{fig:BlockModelFit}, with $b$ = 9.8. The model is used to describe the observation of the physical environment where a single LOS cluster is blocked and when no secondary paths are available. As the TX/RX antenna HPBW increases, the mean signal attenuation decreases since a larger and equal spread of the wavefront impacts the blocker leading to more energy diffracted around the blocker and thus captured by a wider viewing angle at the RX. For very narrow HPBW antennas, the mean signal attenuation will be large since the spread of the transmitted wavefront and viewing angle at the receiver are fully blocked by an obstruction, leading to little-diffracted energy observed at the RX. An interesting note is that there is a limit at which increasing the antenna HPBW will have inconsequential benefits since the LOS path cluster takes such a narrow trajectory between the TX and RX in absence of secondary reflectors and scatterers. As shown in Fig.~\ref{fig:BlockModelFit}, the minimum mean signal attenuation reaches a limit of approximately 10 dB for TX/RX antenna HPBWs of 70$^\circ$ or more. Since only 3 data points were used to model~\eqref{eq:modelFit}, future work will be done to further verify the model. 
\begin{figure}[b!]
	\centering
	\includegraphics[width=0.48\textwidth]{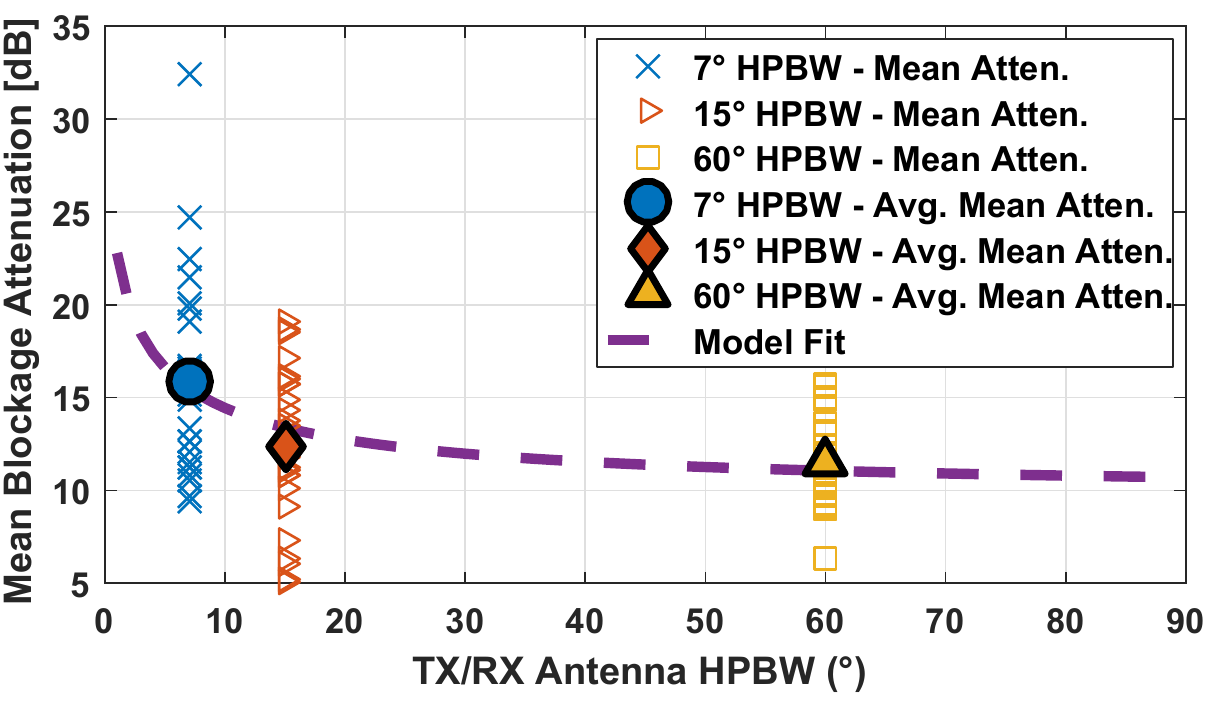}
	\caption{Model fit for~\eqref{eq:modelFit} showing that mean blockage attenuation is inversely related to the TX/RX antenna HPBW.}\label{fig:BlockModelFit}
\end{figure}}

Fade time durations ($t_D$) are shown by the CDFs in Fig.~\ref{fig:fourStateBlockTimeCDF}, where on average the wider beamwidth antennas demonstrated shorter fade durations than narrower beam antennas. Parameters corresponding to the CDF curves best-fit to the measurement data are provided in Table~\ref{tbl:fourStateCDFs}. The larger fade time durations for narrowbeam antennas compared to widebeam antennas is expected since wider beamwidth antennas have larger viewing angles to capture diffracted energy around a pedestrian blocker. Specifically, the 60$^\circ$ HPBW antennas resulted in a mean fade time duration of 321.3 ms, whereas the 7$^\circ$ HPBW antennas demonstrated mean fade durations of 378.0 ms. The larger fade time durations for the narrowbeam antennas might give an indication as to why the signal strength decay and rise rates are slower than for wider beamwidth antennas as observed earlier. Note that the fade time durations for the four-state model are longer than the two-state model since the definitions and thresholds are slightly different in each case (see Sections~\ref{sec:modelTwo} and~\ref{sec:modelFour}). 

Since the durations of fading events are on the order of 100's of milliseconds, mmWave systems must be designed to electrically switch beams on the order of microseconds to find an unblocked reflector or to use TX diversity through a distributed multiple-input and multiple-output (MIMO) architecture where multiple nodes can serve the same user. More specifically, it may be required to implement double-link beam tracking algorithms~\cite{Gao14a} in lieu of single-link beam tracking which is currently implemented in the 802.11ad standard.

The four-state Markov model transition probability rates are provided in Table~\ref{tbl:fourStateProbs} and are useful for creating simple simulations of human blockage events for various antenna HPBW configurations. The CDF parameters provided in Table~\ref{tbl:fourStateCDFs} are also useful for designing PHY and MAC layer protocols, beam switching and beam steering algorithms, or for modeling performance of small cell networks at mmWave frequencies. 

\begin{table}
\centering
\caption{Four-state Markov model transition probability rates for human blockage events at 73.5 GHz}
\label{tbl:fourStateProbs}
\begin{center}
	\scalebox{1}{
		\begin{tabu}{|c|c|c|c|}\hline
			TX/RX HPBW 						& Current State		& Next State	&	$\lambda$ - rate (trans/sec) \\ \specialrule{1.5pt}{0pt}{0pt}
			\multirow{4}{*}{$7^\circ$} 		& unshadowed 		& decaying		&	0.21 \\ \cline{2-4}
											& decaying 			& shadowed		&	7.88 \\ \cline{2-4}
											& shadowed 			& rising		&	7.70 \\ \cline{2-4}
											& rising 			& unshadowed	&	7.67 \\ \hline
			\multirow{4}{*}{$15^\circ$}		& unshadowed 		& decaying		&	0.21 \\ \cline{2-4}
											& decaying 			& shadowed		&	10.49 \\ \cline{2-4}
											& shadowed 			& rising		&	9.79 \\ \cline{2-4}
											& rising 			& unshadowed	&	5.48 \\ \hline
			\multirow{4}{*}{$60^\circ$}		& unshadowed 		& decaying		&	0.18 \\ \cline{2-4}
											& decaying 			& shadowed		&	11.30 \\ \cline{2-4}
											& shadowed 			& rising		&	10.36 \\ \cline{2-4}
											& rising 			& unshadowed	&	6.88 \\ \hline
%			\multirow{2}{*}{$15^\circ$} 	& unshadowed 		& shadowed		&	0.21 \\ \cline{2-4}
%			& shadowed 			& unshadowed	&	3.42 \\ \hline
%			\multirow{2}{*}{$60^\circ$} 	& unshadowed 		& shadowed		&	0.18 \\ \cline{2-4}
%			& shadowed 			& unshadowed	&	3.52 \\ \hline
%			\multirow{2}{*}{All Combined} 	& unshadowed 		& shadowed		&	0.18 \\ \cline{2-4}
%			& shadowed 			& unshadowed	&	3.85 \\ \hline
%			\midrule
%			\multicolumn{5}{c}{\textbf{27 dBi \& 7$^\circ$ Az/El HPBW Antennas}}    \\ \midrule
%							& unshadowed	& decaying	& shadowed	& rising	\\ \midrule
%			unshadowed		& 0.9996		& 0.0004	& 0			& 0			\\
%			decaying		& 0				& 0.9790 	& 0.0210	& 0			\\ 
%			shadowed		& 0				& 0			& 0.9794 	& 0.0206	\\ 
%			rising			& 0.0206		& 0			& 0 		& 0.9794	\\ \midrule
%			\multicolumn{5}{c}{\textbf{20 dBi \& 15$^\circ$ Az/El HPBW Antennas}}    \\ \midrule
%			& unshadowed	& decaying	& shadowed	& rising	\\ \midrule
%			unshadowed		& 0.9993		& 0.0007	& 0			& 0			\\
%			decaying		& 0				& 0.9654 	& 0.0346	& 0			\\ 
%			shadowed		& 0				& 0			& 0.9677 	& 0.0323	\\ 
%			rising			& 0.0181		& 0			& 0 		& 0.9819	\\ \midrule
%			\multicolumn{5}{c}{\textbf{9.1 dBi \& 60$^\circ$ Az/El HPBW Antennas}}    \\ \midrule
%			& unshadowed	& decaying	& shadowed	& rising	\\ \midrule
%			unshadowed		& 0.9994		& 0.0006	& 0			& 0			\\
%			decaying		& 0				& 0.9627 	& 0.0373	& 0			\\ 
%			shadowed		& 0				& 0			& 0.9658 	& 0.0342	\\ 
%			rising			& 0.0227		& 0			& 0 		& 0.9773	\\ \midrule
		\end{tabu}}
	\end{center}
	\vspace{-5mm}
\end{table}

\section{Conclusion}\label{sec:conc}
In this paper, we have demonstrated a mmWave measurement campaign conducted at 73.5 GHz and with 1 GHz of RF bandwidth to study the effects of pedestrian foot-traffic on a peer-to-peer / point-to-point mmWave link in a busy urban scenario. Two simple blockage event models were described from the literature where one used a pre-defined threshold to determine the shadowed and unshadowed states and the second approach models four blockage event states: unshadowed, decaying signal strength, shadowed, and increasing signal strength. The two-state Markov model indicates that shadowing events are longer when using narrowbeam antennas compared to widebeam antennas. The conditional transition probability rates given in Table~\ref{tbl:twoStateProbs} can be used to simulate random blockage events in system and network level simulators for mmWave systems~\cite{Sun17b}. 

The four-state piecewise linear blockage model indicates the rate of decaying signal strength and rising signal strength at the beginning and end of each blockage event by the CDF plots shown in Section~\ref{sec:resultsFour}, where the rates are asymmetrical for the blockage events. The decay and rise rates are likely asymmetrical due to non-symmetrical blockers that caused signal fades when multiple people walked together, or by crossing the LOS path between the TX and RX antennas at non-perpendicular trajectories. The CDFs of the mean signal fade attenuations for the three antenna HPBW pairs demonstrate that average signal fades are inversely related to antenna HPBW such that the 60$^\circ$, 15$^\circ$, and 7$^\circ$ TX/RX HPBW antenna pairs had monotonically increasing mean signal attenuations of 11.5 dB, 12.4 dB, and 15.8 dB, respectively. A simple model for mean signal attenuation as a function of antenna HPBW is introduced for this physical observation. The observation that fade depth is a function of antenna beamwidth is similar to other work that showed path loss, delay spread, and azimuth spreads are functions of antenna beamwidth and frequency~\cite{Rap15b}. Furthermore, previous studies reported similar observations with mean signal attenuations between 6 dB and 18 dB~\cite{Jacob10b} at 60 GHz. On average, blockage events were longer for narrower beamwidth antennas since they have a much smaller viewing angle to capture energy as compared to wider HPBW antennas, as shown by the 378.0 ms mean fade time duration for 7$^\circ$ HPBW antennas compared to 321.3 ms for 60$^\circ$ HPBW antennas. A remarkable result is that the median fade time durations are approximately 255 ms and 327 ms for the two-state and four-state models, respectively, no matter what the TX and RX antenna beamwidths are, which may be a result of only detecting a LOS path cluster over all measurements. The fade durations reported here are in agreement with earlier studies at 60 GHz that were on the order of 300 ms to 500 ms~\cite{Jacob10b,Collonge04a}. 

The blockage events observed here for peer-to-peer networks in a dense urban crowded area are important for designing mmWave communications systems that rely on electrically steerable antenna systems to consistently deliver seamless connectivity and high-throughput in the presence of pedestrians and rapid fading. As the technology for electrically-steerable and reconfigurable antenna arrays matures, future studies will be necessary for understanding the performance of switching antenna beams at microsecond speeds to find reflectors and scatterers to maintain sufficient SNR when the dominant path between the TX and RX is heavily obstructed.

\bibliography{MacCartney_Bibv6}

\begin{thebibliography}{10}
\providecommand{\url}[1]{#1}
\csname url@samestyle\endcsname
\providecommand{\newblock}{\relax}
\providecommand{\bibinfo}[2]{#2}
\providecommand{\BIBentrySTDinterwordspacing}{\spaceskip=0pt\relax}
\providecommand{\BIBentryALTinterwordstretchfactor}{4}
\providecommand{\BIBentryALTinterwordspacing}{\spaceskip=\fontdimen2\font plus
\BIBentryALTinterwordstretchfactor\fontdimen3\font minus
  \fontdimen4\font\relax}
\providecommand{\BIBforeignlanguage}[2]{{%
\expandafter\ifx\csname l@#1\endcsname\relax
\typeout{** WARNING: IEEEtran.bst: No hyphenation pattern has been}%
\typeout{** loaded for the language `#1'. Using the pattern for}%
\typeout{** the default language instead.}%
\else
\language=\csname l@#1\endcsname
\fi
#2}}
\providecommand{\BIBdecl}{\relax}
\BIBdecl

\bibitem{Rap13a}
T.~S. Rappaport \emph{et~al.}, ``{Millimeter Wave Mobile Communications for
  {5G} Cellular: It Will Work!}'' \emph{IEEE Access}, vol.~1, pp. 335--349, May
  2013.

\bibitem{Boccardi14a}
F.~Boccardi \emph{et~al.}, ``Five disruptive technology directions for {5G},''
  \emph{IEEE Communications Magazine}, vol.~52, no.~2, pp. 74--80, Feb. 2014.

\bibitem{3GPP.38.901}
3GPP, ``Technical specification group radio access network; study on channel
  model for frequencies from 0.5 to 100 {GHz (Release 14)},'' 3rd Generation
  Partnership Project (3GPP), TR 38.901 V1.0.1, Mar. 2017.

\bibitem{A5GCM15}
\BIBentryALTinterwordspacing
{Aalto University, AT\&T, BUPT, CMCC, Ericsson, Huawei, Intel, KT Corporation,
  Nokia, NTT DOCOMO, New York University, Qualcomm, Samsung, University of
  Bristol, and University of Southern California}, ``{5G} channel model for
  bands up to 100 {GHz},'' 2016, Oct. 21. [Online]. Available:
  \url{http://www.5gworkshops.com/5GCM.html}
\BIBentrySTDinterwordspacing

\bibitem{Sun16b}
S.~Sun \emph{et~al.}, ``Investigation of prediction accuracy, sensitivity, and
  parameter stability of large-scale propagation path loss models for {5G}
  wireless communications ({Invited Paper}),'' \emph{IEEE Transactions on
  Vehicular Technology}, vol.~65, no.~5, pp. 2843--2860, May 2016.

\bibitem{Sun17b}
S.~Sun, G.~R. {MacCartney, Jr.}, and T.~S. Rappaport, ``A novel millimeter-wave
  channel simulator and applications for {5G} wireless communications,'' in
  \emph{2017 IEEE International Conference on Communications (ICC)}, May 2017,
  pp. 1--7.

\bibitem{Sun17a}
S.~Sun \emph{et~al.}, ``Millimeter wave small-scale spatial statistics in an
  urban microcell scenario,'' in \emph{2017 IEEE International Conference on
  Communications (ICC)}, May 2017, pp. 1--7.

\bibitem{Eliasi15a}
P.~A. Eliasi and S.~Rangan, ``Stochastic dynamic channel models for millimeter
  cellular systems,'' in \emph{2015 IEEE 6th International Workshop on
  Computational Advances in Multi-Sensor Adaptive Processing ({CAMSAP})}, Dec.
  2015, pp. 209--212.

\bibitem{Zhang16b}
M.~Zhang \emph{et~al.}, ``Transport layer performance in {5G mmWave}
  cellular,'' in \emph{2016 IEEE Conference on Computer Communications
  Workshops (INFOCOM WKSHPS)}, Apr. 2016, pp. 730--735.

\bibitem{Collonge04a}
S.~Collonge, G.~Zaharia, and G.~{El Zein}, ``Influence of the human activity on
  wide-band characteristics of the 60 {GHz} indoor radio channel,'' \emph{IEEE
  Transactions on Wireless Communications}, vol.~3, no.~6, pp. 2396--2406, Nov.
  2004.

\bibitem{Jacob10b}
M.~Jacob, C.~Mbianke, and T.~K{\"u}rner, ``A dynamic 60 {GHz} radio channel
  model for system level simulations with {MAC} protocols for {IEEE
  802.11ad},'' in \emph{IEEE International Symposium on Consumer Electronics
  (ISCE 2010)}, June 2010, pp. 1--5.

\bibitem{Jacob09d}
------, ``Human body blockage - guidelines for {TGad MAC} development,'' doc.:
  IEEE 802.11-09/1169r0, Nov. 2009.

\bibitem{Mac16a}
G.~R. {MacCartney, Jr.} \emph{et~al.}, ``Millimeter-wave human blockage at 73
  {GHz} with a simple double knife-edge diffraction model and extension for
  directional antennas,'' in \emph{2016 IEEE 84th Vehicular Technology
  Conference (VTC2016-Fall)}, Sept. 2016, pp. 1--6.

\bibitem{80211ad10a}
A.~Maltsev \emph{et~al.}, ``{Channel Models for 60 GHz WLAN Systems},'' doc.:
  IEEE 802.11-09/0334r8, May 2010.

\bibitem{Weiler14a}
R.~J. Weiler \emph{et~al.}, ``Measuring the busy urban 60 {GHz} outdoor access
  radio channel,'' in \emph{2014 IEEE International Conference on
  Ultra-WideBand (ICUWB)}, Sept. 2014, pp. 166--170.

\bibitem{Weiler16a}
------, ``Environment induced shadowing of urban millimeter-wave access
  links,'' \emph{IEEE Wireless Communications Letters}, vol.~5, no.~4, pp.
  440--443, Aug. 2016.

\bibitem{Peter12a}
M.~Peter \emph{et~al.}, ``Analyzing human body shadowing at 60 {GHz}:
  Systematic wideband {MIMO} measurements and modeling approaches,'' in
  \emph{2012 6th European Conference on Antennas and Propagation (EuCAP)}, Mar.
  2012, pp. 468--472.

\bibitem{Kunisch08a}
J.~Kunisch and J.~Pamp, ``Ultra-wideband double vertical knife-edge model for
  obstruction of a ray by a person,'' in \emph{2008 IEEE International
  Conference on Ultra-Wideband}, vol.~2, Sept. 2008, pp. 17--20.

\bibitem{Malik07a}
W.~Q. Malik, B.~Allen, and D.~J. Edwards, ``Impact of bandwidth on small-scale
  fade depth,'' in \emph{IEEE GLOBECOM 2007 - IEEE Global Telecommunications
  Conference}, Nov. 2007, pp. 3837--3841.

\bibitem{Holtzman94a}
J.~M. Holtzman and L.~M.~A. Jalloul, ``Rayleigh fading effect reduction with
  wideband {DS/CDMA} signals,'' \emph{IEEE Transactions on Communications},
  vol.~42, no. 234, pp. 1012--1016, Feb. 1994.

\bibitem{Mac17a}
G.~R. {MacCartney, Jr.} and T.~S. Rappaport, ``A flexible millimeter-wave
  channel sounder with absolute timing,'' \emph{IEEE Journal on Selected Areas
  in Communications}, vol.~35, no.~6, pp. 1402--1418, June 2017.

\bibitem{Kashiwagi10a}
I.~Kashiwagi, T.~Taga, and T.~Imai, ``Time-varying path-shadowing model for
  indoor populated environments,'' \emph{IEEE Transactions on Vehicular
  Technology}, vol.~59, no.~1, pp. 16--28, Jan. 2010.

\bibitem{Dehnie07a}
S.~Dehnie, ``Markov chain approximation of rayeleigh fading channel,'' in
  \emph{2007 IEEE International Conference on Signal Processing and
  Communications}, Nov. 2007, pp. 1311--1314.

\bibitem{Jacob10a}
M.~Jacob \emph{et~al.}, ``Modeling the dynamical human blockage for 60 {GHz
  WLAN} channel models,'' in \emph{\normalfont{doc.: IEEE 802.11-10/0090r0}},
  Jan. 2010.

\bibitem{Rap15b}
T.~S. Rappaport \emph{et~al.}, ``Wideband millimeter-wave propagation
  measurements and channel models for future wireless communication system
  design ({Invited Paper}),'' \emph{IEEE Transactions on Communications},
  vol.~63, no.~9, pp. 3029--3056, Sept. 2015.

\bibitem{Singh09a}
S.~Singh \emph{et~al.}, ``Blockage and directivity in 60 {GHz} wireless
  personal area networks: from cross-layer model to multihop {MAC} design,''
  \emph{IEEE Journal on Selected Areas in Communications}, vol.~27, no.~8, pp.
  1400--1413, Oct. 2009.

\bibitem{Gao14a}
B.~Gao \emph{et~al.}, ``Double-link beam tracking against human blockage and
  device mobility for 60-{GHz WLAN},'' in \emph{2014 IEEE Wireless
  Communications and Networking Conference (WCNC)}, Apr. 2014, pp. 323--328.

\end{thebibliography}
\bibliographystyle{IEEEtran}
\end{document}